\documentstyle[11pt]{article}
\newtheorem{theorem}{\bf Theorem}
\newtheorem{lemma}{\bf Lemma}
\newtheorem{corollary}{\bf Corollary}

\newtheorem{definition}{\bf Definition}

\newenvironment{proof}{\par \bf Proof. \rm}{$\Box$ \vspace{1ex}}

\begin{document}
\date{}

\title{\Large\bf Reversible Simulation of Irreversible Computation by Pebble Games
}
\author{
Ming Li\thanks{Supported in part by
the NSERC Operating Grant OGP0046506, ITRC, a CGAT grant, and the
Steacie Fellowship. Current address: Department of Computer Science,
City University of Hong Kong, Kowloon, Hong Kong. On sabbatical leave
from: Department of Computer Science, University of Waterloo,
Waterloo, Ont. N2L 3G1, Canada. E-mail: mli@cs.cityu.edu.hk;
mli@math.uwaterloo.ca} 
\and
John Tromp\thanks{Partially
supported by the European Union
through NeuroCOLT ESPRIT Working Group Nr. 8556,
and by  NWO through NFI Project ALADDIN under Contract
number NF 62-376 and
NSERC under
International Scientific Exchange Award ISE0125663. Affiliations
are CWI and the University of Amsterdam.
Address: CWI, Kruislaan 413, 1098 SJ Amsterdam, The Netherlands; tromp@cwi.nl}
\and 
Paul Vit\'{a}nyi\thanks{Partially
supported by the European Union
through NeuroCOLT ESPRIT Working Group Nr. 8556,
and by  NWO through NFI Project ALADDIN under Contract
number NF 62-376 and
NSERC under
International Scientific Exchange Award ISE0125663. Affiliations
are CWI and the University of Amsterdam.
Address: CWI, Kruislaan 413, 1098 SJ Amsterdam, The Netherlands; paulv@cwi.nl}
}

\maketitle
\begin{abstract}
{\em 
Reversible simulation of irreversible algorithms is analyzed
in the stylized form of a `reversible' pebble game.
%While such
%simulations incur little overhead in additional
%computation time, they use a large
%amount of additional memory space during the computation.
The reacheable reversible
simulation instantaneous descriptions (pebble configurations) 
are characterized completely. As a corollary we obtain
the reversible simulation by Bennett and
that among all simulations that can be modelled by
the pebble game, Bennett's simulation is optimal in that it
uses the least auxiliary space for the greatest number of
simulated steps. One can reduce the auxiliary storage overhead
incurred by the reversible simulation at the cost of allowing
limited erasing leading to an irreversibility-space tradeoff.
We show that in this resource-bounded setting 
the limited erasing needs to be performed at precise
instants during the simulation. We show that the
reversible simulation can be modified 
so that it is applicable also when the simulated
computation time is unknown.
}
\end{abstract}

\section{Introduction}
The ultimate limits of miniaturization of computing
devices, and therefore the speed of computation,
are constrained
by  the increasing
density of switching elements in the device. 
Linear speed up
by shortening interconnects on a two-dimensional device
is attended by a cubing of
dissipated energy per unit area per second. Namely, we square
the number of switching elements
per area unit and linearly increase the number of switching
events per switch per time unit. In the long run, the attending
energy dissipation on this scale cannot be compensated for by cooling.
Ignoring architectural improvements,
reduction of the energy dissipation
per elementary computation step therefore determines future advances
in computing power.

%Since 1940 the
%dissipated energy per bit operation in a computing
%device has---with remarkable regularity---decreased
%by roughly one order of magnitude (tenfold) every five years \cite{Ke88,La88}.
%Extrapolation of current trends
%shows that the energy dissipation per
%binary logic operation needs to be reduced below $kT$
%(thermal noise)
%within 20 years. Here $k$ is Boltzmann's constant and $T$
%the absolute temperature in degrees Kelvin,
%so that $kT \approx 3 \times 10^{-21}$
%Joule at room temperature. Even at $kT$ level, 
%a future device containing $10^{12}$ gates
%%in a cubic centimeter %who cares:-?
%operating at a gigahertz dissipates about 3 Joule/second. 
%(This on the assumption that all gates are switched all the time.)
%For thermodynamic reasons,
%cooling the operating temperature of such a computing device
%to almost absolute zero (to get $kT$ down) must
%dissipate at least as much energy in the cooling as it saves
%for the computing \cite{Me93}.
%
%Considerations of thermodynamics of
%computation started 
%in the early fifties. 
J. von Neumann reputedly thought
that a computer operating at temperature $T$ must
dissipate at least $k T \ln 2$ Joule per elementary bit 
operation \cite{Bu}.
R. Landauer~\cite{La61}
demonstrated
that it is only the `logically 
irreversible' operations in a physical computer
that are required to dissipate energy by generating a
corresponding amount of entropy for each bit of information
that gets irreversibly erased. As a consequence,
any arbitrarily
large reversible computation can be performed on an appropriate
physical device using only
one unit of physical energy {\em in principle}.  

\section{Reversible Turing Machines}
Currently, we are used to design
computational procedures containing irreversible operations.
To perform the intended computations
 without energy dissipation the related computation procedures need to become
completely reversible.
Fortunately, all irreversible computations 
can be simulated in a reversible manner \cite{Le63,Be73}.
All known reversible simulations do not change the computation time
significantly, but do require considerable amounts
of auxiliary memory space.
In this type of simulation one
needs to save on space; time is already almost optimal.

Consider the standard model of Turing machine.
The elementary operations are rules
in quadruple format $(p,s,a,q)$ meaning that
if the finite control is in state $p$ and the machine
scans tape symbol $s$, then the machine performs action $a$
and subsequently the finite control enters state $q$.
Such an action $a$ consists of either printing a symbol $s'$
in the tape square under scan, or moving the scanning head
one tape square left,
right or not at all.

Quadruples are said
to {\em overlap in domain} if they cause the machine in the same state
and scanning the same symbol to perform different actions.
A {\em deterministic Turing machine} is defined as a Turing machine
with quadruples no two of which overlap in domain.
 
Now consider the special format (deterministic) Turing
machines using quadruples of two
types: {\em read/write} quadruples and {\em move} quadruples.
A read/write quadruple $(p,a,b,q)$ causes the machine in state
$p$ scanning tape symbol $a$ to write symbol $b$ and enter state $q$.
A move quadruple $(p,\ast,\sigma ,q)$ causes the machine
in state $p$ to move its tape head by $\sigma \in \{-1,0,+1\}$
squares and enter state $q$,
oblivious to the particular symbol in the currently scanned tape square.
(Here `$-1$' means `one square left', `$0$' means `no move'
and `$+1$' means `one square right'.) Quadruples are said
to {\em overlap in range} if they cause the machine to enter
the same state and either both write the same symbol or
(at least) one of them moves the head. Said differently,
quadruples that enter the same state overlap in range
unless they write different symbols.
A {\em reversible Turing machine} is a deterministic Turing machine
with quadruples no two of which overlap in range.
A $k$-tape reversible Turing machine uses $(2k+2)$ tuples
which, for each tape separately, select a read/write or move on that
tape. Moreover, any two tuples can be restricted to some single tape
where they don't overlap in range.
 
To show that each partial recursive function can be computed
by a reversible Turing machine one can proceed as follows.
Take the standard irreversible Turing machine computing that function.
We modify it by
adding an auxiliary storage tape called the `history tape'.
The quadruple rules are extended to 6-tuples to additionally
manipulate the history tape.
To be able to reversibly undo (retrace)
the computation deterministically, the new 6-tuple
rules have the effect that the machine keeps a record
on the auxiliary history tape consisting of
the sequence of quadruples executed on the original tape.
Reversibly undoing a computation
entails also erasing the record
of its execution from the history tape.

This notion of reversible computation means
that only $1:1$ recursive functions can be computed.
To reversibly simulate $t$ steps of an
irreversible computation from $x$ to $f(x)$
one reversibly  computes from input $x$
to output $\langle x, f(x) \rangle$.
%Say, this takes $t' = O(t)$ time.
%Since this reversible simulation at some time instant
%has to record the entire
%history of the irreversible computation, its space use increases
%linear with the number of simulated steps $t$. That is,
%if the simulated irreversible computation uses $s$ space, then
%for some constant $c > 1$ the simulation uses
%$t'\approx c+ct$ time and $s'\approx c + c(s+t)$ space.
%After computing from $x$ to $f(x)$
%the machine reversibly copies $f(x)$, reversibly undoes
%the computation from  $x$ to $f(x)$ erasing its history tape
%in the process, and ends with one copy of $x$ and
%one copy of $f(x)$ in the format $\langle x, f(x) \rangle$
%and otherwise empty tapes.

\section{Reversible Programming}

Reversible Turing machines or other reversible computers
will require special reversible programs. One feature of such
programs is that they should be executable when read from bottom
to top as well as when read from top to bottom. Examples are
the programs $F(\cdot)$ and $A(\cdot)$ we show in the later sections.
In general, writing reversible programs will be difficult.
However, given a general reversible simulation of irreversible computation,
one can simply write an oldfashioned 
irreversible program in an irreversible programming language,
and subsequently simulate it reversibly. Below we address
the question of parsimonious resource-bounded reversible simulations.

In terms of real computers, eventually there will be reversible
computer architectures running reversibly programmed compilers.
Such a compiler receives an irreversible program as input and
 reversibly compiles
it to a reversible program. Subsequently, the reversible program is
executed reversibly.

There is a decisive difference between reversible circuits and
reversible special purpose computers on the one hand, and
reversible universal computers on the other hand.
While one can design a special-purpose reversible version for each particular
reversible circuit using reversible universal gates, such a method
does not suffice to execute an arbitrary irreversible
program on a fixed universal reversible computer architecture. 
In this paper we
are interested the latter question:
the design of methods which take any irreversible program
and make it run in a reversible manner on a fixed universal reversible machine.

\section{Related Work}
The reversible simulation in \cite{Be73} of $T$ steps of an
irreversible computation from $x$ to $f(x)$
reversibly  computes from input $x$ 
to output $\langle x, f(x) \rangle$
in $T' = O(T)$ time.
%The number of
%irreversibly provided and erased bits is
%therefore upper bounded as $E^{O(T)} (x,f(x)) \leq 2|p|+|x|$. 
%For erasure, with $f(x)= \epsilon$ is a function which has
%a very short irreversible program $p$ of $O(1)$ bits,
 %we obtain $E^{O(T)}(x, \epsilon)
%\leq |x| + O(1)$.
However, since this reversible simulation at some time instant
has to record the entire
history of the irreversible computation, its space use increases
linearly with the number of simulated steps $T$. That is,
if the simulated irreversible computation uses $S$ space, then
for some constant $c > 1$ the simulation uses 
$T'\approx c+cT$ time and $S'\approx c + c(S+T)$ space.
This can be an unacceptable amount of space for many practically 
useful computations. 

The question arises whether one can reduce
the amount of auxiliary space needed by the simulation by a
more clever simulation method or
by allowing limited amounts of irreversibility.

In \cite{Be89} another elegant simulation technique is devised
reducing the auxiliary storage space.
This simulation does not save the entire history of the irreversible
computation but it breaks up the simulated computation
into segments of about $S$ steps
 and saves in a hierarchical manner {\em checkpoints}
consisting of complete instantaneous descriptions of the
simulated machine (entire tape contents, tape heads positions,
state of the finite control). After a later checkpoint is
reached and saved, the simulating machine reversibly
undoes its intermediate computation, reversibly erasing
the intermediate history and reversibly canceling the previously
saved checkpoint. Subsequently, the computation is resumed from
the new checkpoint onwards. 

The reversible computation simulates $k^n$ segments of length $m$
of irreversible
computation in $(2k-1)^n$ segments of length $\Theta (m+S)$
of reversible computation using
$n(k-1)+1$ checkpoint registers using $\Theta (m+S)$ 
space each, for each $k,n,m$. 

This way it is established that there are various trade-offs 
possible in time-space in between $T'= \Theta (T)$ and
$S' = \Theta (TS)$ at one extreme ($k=1, m=T, n=1$) and (with the corrections
of \cite{LeSh90})
$T' = \Theta (T^{1+\epsilon}/S^{\epsilon} )$ 
and $S'= \Theta ( c(\epsilon) S(1+ \log T/S))$
with $c(\epsilon)= \epsilon 2^{1/\epsilon}$
for each $\epsilon > 0$, 
using always the same simulation method but with different
parameters $k,n$ where $\epsilon = \log_k (2k-1)$ and $m = \Theta (S)$. 
Typically, for $k=2$ we have $\epsilon = \log 3$.
Since for $T > 2^S$ the machine goes into
a computational loop, we always have $S \leq \log T$.
Therefore, it follows from Bennett's simulation
that  each irreversible Turing machine 
using space $S$ can be simulated by a reversible machine 
using space $S^2$ in polynomial time.

In \cite{LiVi96} two of us proposed a quantitative study of
exchanges of computing resources such as time and space
for number of irreversible operations which we believe will be
relevant for the physics of future computation devices.

There, we analyzed
the advantage of adding limited irreversibility
to an otherwise reversible simulation of conventional
irreversible computations. This may be
of some practical relevance for adiabatic computing.
Our point of departure is the
general method of Bennett \cite{Be89} to
reversibly simulate irreversible algorithms
in the stylized form of a pebble game. 
While such 
reversible simulations incur little overhead in additional
computation time, they use an unacceptable
amount of additional memory space during the computation.
We showed that among all simulations which can be modelled by
the pebble game, Bennett's simulation is optimal in that it
uses the least auxilliary space for the greatest number of
simulated steps. 

In that paper
we {\em conjectured} that {\em all} reversible simulations
of an irreversible computation can
essentially be represented as the pebble game defined below,
and that consequently the lower bound of Corollary~\ref{lem.pebble}
applies to all reversible simulations of irreversible
computations. Contradicting this conjecture,
\cite{LMT97} have shown that 
there exists a linear space reversible simulation of an
irreversible computation of the following type.
The original irreversible
computation should have a reversible linear space constructable space bound,
and the reversible simulation of a given irreversible computation---when run 
in reverse---cycles through the set of {\em all} initial configurations
that cause the irreversible computation to terminate in the target
final configuration
rather than homing in on
the {\em single} initial configuration of the irreversible computation.
The reversible simulation time can be exponential in the
irreversible computation time.
This reversible simulation of the irreversible
computation $x \mapsto f(x)$ actually simulates {\em all}
irreversible computations $y \mapsto f(y)$ such that $f(y)=f(x)$.

Instructed by \cite{LMT97} we refine the conjecture
to: ``Polynomial time reversible simulations
of an irreversible computation can
essentially be represented as the pebble game defined below,
and consequently the lower bound of Corollary~\ref{lem.pebble}
applies to all polynomial time reversible simulations of irreversible
computations.'' In fact, for ``polynomial time'' one should
actually think of, say, at most ``square time.'' What one wants is
that the reversible simulation is {\em feasible} which means that
the both the simulation time overhead and the simulation space overhead 
do not increase too much.
If this conjecture is true then
the trade-offs below turn into a space-irreversibility
hierarchy for polynomial time computations.

\section{Outline Current Work}

We improve and extend the reversible simulation 
work in \cite{Be89,LiVi96}.
First, we note that it is possible to improve the situation
by reversibly simulating only irreversible steps. 
Call a quadruple of a Turing machine {\em irreversible} if
its range overlaps with the range of another quadruple.
A step of the computation is {\em irreversible} if
it uses an irreversible quadruple. Let the number of
irreversible steps in a $T$ step computation be denoted by $I$.
Clearly, $I \leq T$. Then, Bennett's simulation results
hold with $T$ in the auxiliary space use replaced by $I$.
In particular, $S'=O(S\log I)$. In many computations,
$I$ may be much smaller than $T$. There arises the problem
of estimating the number of irreversible  steps in a
computation. (More complicatedly, one could extend the notion
of irreversible step to those steps which can be reversed on local information
alone. In some cases this is possible
 even when the used quadruple itself was irreversible.)

In the sequel we first define the reversible pebble game.
We then completely characterize the realizable
pebble configurations of the reversible pebble games.
That is, we completely characterize the reacheable instantaneous
descriptions of a Turing machine reversibly simulating an
irreversible computation. As corollaries we obtain Bennett's
simulation result, \cite{Be89}, the impossibility result
and the irreversibility tradeoff in \cite{LiVi96}.
Subsequently, we show that for such a tradeoff to work
the limited ireversible actions have to take place at precise
times during the reversible simulation, and cannot be delayed
to be executed all together at the end of the computation (as is
possible in computations without time or space resource bounds).
Finally, in all such reversible simulations it is assumed that
the number of steps to be simulated is known in advance and used
to construct the simulation (for that number of steps).
We show how to reversibly simulate an irreversible computation of
unknown computing time, using the same order of magnitude of simulation time.

\section{Reversible Simulation}
Analyzing the
simulation method of \cite{Be89} 
shows that it is essentially no better than
the simple \cite{Be73} simulation in terms of time
versus irreversible erasure trade-off. 
Extra irreversible erasing 
can reduce the simulation time of the former method to $\Theta (T)$,
but the `simple' method has $\Theta (T)$ simulation
time without irreversible
erasures anyway, but at the cost of large space consumption. 
Therefore, it is crucial 
to decrease the extra space
required for the pure reversible simulation without 
increasing time if possible,
and in any case further reduce the extra space 
at the cost of limited numbers of irreversible erasures.

Since no general reversible simulation
of an irreversible computation, better than the above one, is known,
and it seems likely that each proposed method must
have similar history preserving features,
analysis of this particular style of simulation is
expected to give results with more general validity.
We establish lower bounds on space use and upper bounds
on space versus irreversible erasure trade-offs.

To analyze such trade-offs we use Bennett's
brief suggestion in \cite{Be89}
that a reversible simulation can be modelled by the following
`reversible' pebble game. Let $G$ be a linear list of
nodes $\{1,2, \ldots , T_G \}$.
We define a {\em pebble game} on $G$ as follows. The game
proceeds in a discrete sequence of steps of a single {\em player}.
There
are $n$ pebbles which can be put on nodes of $G$.
At any time the set of pebbles is divided in 
pebbles on nodes of $G$ and the remaining pebbles which are called
{\em free} pebbles. At each step either an existing
 free pebble can be put
on a node of $G$ (and is thus removed from the free pebble pool)
 or be removed from a node of $G$ (and is added to the
free pebble pool).
Initially $G$ is unpebbled and there is a pool of free pebbles.
The game is played according to the following rule:

\begin{description}
\item[Reversible Pebble Rule:]
If node $i$ is occupied by a pebble, then one may either
place a free pebble on node $i+1$ (if it was not occupied before), or
remove the pebble from node $i+1$.
\end{description}

We assume an extra initial node $0$ permanently 
occupied by an extra, fixed pebble,
so that node $1$ may be (un)pebbled at will.
This pebble game is inspired by the method of simulating irreversible Turing
Machines on reversible ones in a space efficient manner. The placement
of a pebble corresponds to checkpointing the current state of the irreversible
computation, while the removal of a pebble corresponds to reversibly erasing
a checkpoint. Our main interest is in determining the number of pebbles $k$
needed to pebble a given node $i$.

The maximum number $n$ of pebbles 
which are simultaneously on $G$ 
at any one time in the game gives the space complexity
$nS$ of the simulation. If one deletes a pebble not following
the above rules, then this means a block of bits of size $S$ is 
erased irreversibly. The limitation to
Bennett's simulation is in fact space, rather than time.
When space is limited, we may not have enough place to store garbage,
and these garbage bits will have to be irreversibly erased.
We establish a tight lower bound for {\em any}
strategy for the pebble game in order to obtain
a space-irreversibility trade-off.

\section{Reachable Pebble Configurations}
\label{reachable}

We describe the idea of Bennett's simulation \cite{Be89}.
Given that some node $s$ is pebbled, and that at least
$n$ free pebbles are available, the task of pebbling nodes $s+1,\ldots,
s+2^n-1$ can be seen to reduce to the task of first pebbling nodes
$s+1,\ldots,s+2^{n-1}-1$ using $n-1$ free pebbles, then placing a free pebble
on node $s+2^{n-1}$, then unpebbling nodes $s+1,\ldots,s+2^{n-1}-1$ to
retrieve our $n-1$ pebbles, and finally pebbling nodes
$s+2^{n-1}+1,\ldots,s+2^n-1$ using these pebbles. By symmetry,
an analogous reduction works for the task of unpebbling nodes $s+1,\ldots,
s+2^n-1$ with $n$ free pebbles. The following two mutually recursive
procedures implement this scheme; their correctness follows by
straightforward induction.

\begin{tabbing}
pe\=bble($s,n$) \\
\{ \\
\> if ($n=0$) return; \\
\> t = s + $2^{n-1}$; \\
\> pebble($s,n-1$); \\
\> put a free pebble on node $t$ \\
\> unpebble($s,n-1$) \\
\> pebble($t,n-1$); \\
\} \\
 \\
unpebble(s,n) \\
\{ \\
\> if ($n=0$) return; \\
\> t = s + $2^{n-1}$; \\
\> unpebble($t,n-1$); \\
\> pebble($s,n-1$) \\
\> remove the pebble from node $t$ \\
\> unpebble($s,n-1$); \\
\}
\end{tabbing}

The difficult part is showing that this method is optimal.
It turns out that characterizing the maximum node that can be pebbled with
a given number of pebbles is best done by completely characterizing
what pebble configurations are realizable.
First we need to introduce some helpful notions.

In a given pebble configuration with $f$ free pebbles,
a placed pebble is called {\em available} if there is another pebble at most
$2^f$ positions to its left ($0$ being the leftmost node).
According to the above procedures, an available pebble can be removed
with the use of the free pebbles. For convenience we imagine this as
a single big step in our game.

Call a pebble configuration {\em weakly solvable} if there is a way
of repeatedly removing an available pebble until all are free.
Note that such configurations are necessarily realizable, since the removal
process can be run in reverse to recreate the original configuration.
Call a pebble configuration {\em strongly solvable} if all ways of repeatedly
removing an available pebble lead to all being free. Obviously any
strongly solvable configuration is also weakly solvable.

The starting configuration is obviously both weakly and strongly
solvable. How does the single rule of the game affect solvability?
Clearly, adding a pebble to a weakly solvable configuation yields
another weakly solvable configuation, while removing a pebble from a strongly
solvable configuation yields another strongly solvable configuation.
It is not clear if removing a pebble from a weakly solvable configuation
yields another one. If such is the case then we may conclude that all
realizable configurations are weakly solvable and hence the two classes
coincide. This is exactly what the next theorem shows.

\begin{theorem}
Every weakly solvable configuration is strongly solvable.
\end{theorem}

\begin{proof}
Let $f$ be the number of free pebbles in a weakly solvable
configuration.
%It suffices to show that removal of any available pebble yields another
%weakly solvable configuration.
Number the placed pebbles $f,f+1,\ldots,n-1$ according
to their order of removal. It is given that, for all $i$,
pebble $i$ has a higher-numbered pebble at most $2^i$ positions
to its left (number the fixed pebble at $0$ infinity). Suppose 
a pebble of rank $g>f$ pebble is also available. 
It suffices to show that if pebble $g$ is removed first, then pebbles
$f,f+1,\ldots,g-1$ are still available when their turn comes.
Suppose pebble $j$ finds pebble $g$ at most $2^j$ places to its left
(otherwise it will still be available for sure).
Then after removal of pebbles $g,f,f+1,\ldots,j-1$, it will still find
a higher-numbered pebble at most $2^j + 2^f + 2^{f+1} + \cdots + 2^{j-1}
\leq 2^{j+1}$ places to its left, thus making it available given the extra
now free pebble $g$.
\end{proof}

\begin{corollary}
A configuration with $f$ free pebbles
is realizable if and only if its placed pebbles can be
numbered $f,f+1,\ldots,n-1$ such that
pebble $i$ has a higher-numbered pebble at most $2^i$ positions
to its left.
\end{corollary}

\begin{corollary}\label{lem.bennett}\label{lem.pebble}
The maximum reachable node with $n$ pebbles is $\sum_{i=0}^{n-1} 2^i = 2^n-1$.
\end{corollary}

Moreover, if pebble($s,n$) takes $t(n)$ steps
we find $t(0) = 1$ and
$t(n)=3 t(n-1) + 1 = (3^{n+1}-1)/2$. That is, the number
of steps $T_G'$ of a winning play of a pebble game
of size $T_G=2^n-1$ is $T_G' \approx 1.5 3^n$, that is,
$T_G' \approx T_G^{\log 3}$.

The simulation given in \cite{Be89} follows the rules
of the pebble game of length $T_G = 2^n-1$ with $n$ pebbles above.
A winning
strategy for a game of length $T_G$ using $n$ pebbles
corresponds with reversibly simulating $T_G$ segments of $S$
steps of an irreversible computation using $S$
space such that the reversible simulator
uses $T' \approx ST'_G \approx ST_G^{\log 3}$ steps 
and total space $S'=nS$. The space $S'$ corresponds
to the maximal number of pebbles on $G$
at any time during the game.  The placement or removal of a
pebble in the game corresponds to the reversible
copying or reversible cancelation of a `checkpoint'
consisting of the entire instantaneous description of size $S$
(work tape contents, location of heads, state of finite
control) of the simulated irreversible machine.
The total time $T_GS$ used by the irreversible computation
is broken up in segments of size $S$ so that the reversible
copying and canceling of a checkpoints takes about the same
number of steps as the computation segments in between
checkpoints.
\footnote{If we are to account for the permanent pebble on node $0$,
we get that the simulation uses $n+1$
pebbles for a pebble game with $n$ pebbles of length $T_G+1$.
The simulation uses $n+1 =S'/S$ pebbles for
a simulated number of $S(T_G+1)$ steps of the irreversible
computation.}

We can now formulate a trade-off between space used
by a polynomial time reversible computation and irreversible
erasures. First we show that allowing a limited
amount of erasure in an otherwise
reversible computation means that
we can get by with less work space.
Therefore, we define an {\em $m$-erasure} pebble game as
the pebble game above but with the additional rule

\begin{itemize}
\item
In at most $m$ steps
the player can 
remove a pebble from any node $i > 1$ without
node $i-1$ being pebbled at the time.
\end{itemize}
 
An $m$-erasure pebble game corresponds with an otherwise
reversible computation using $mS$ irreversible bit erasures,
where $S$ is the space used by the irreversible computation
being simulated.

\begin{lemma}\label{lem.erasure}
There is a winning strategy with $n+2$ pebbles
and $m-1$ erasures for pebble games $G$
with $T_G= m2^n$, for all $m \geq 1$.
\end{lemma}
\begin{proof}
The strategy is to use 2 pebbles as springboards that are alternately
placed $2^n$ in front of each other using the remaining $n$ pebbles
to bridge the distance. The most backward springboard can be erased
from its old position once all $n$ pebbles are cleared from the space
between it and the front springboard.
%using $n$ pebbles without
%erasures (as in Corollary~\ref{lem.bennett}),
%put the $n$th pebble in front, and invert 
%the advancement process to free all the
%pebbles in the block. The last remaining
%pebble has no predecessor and needs to be
%irreversibly erased except in the initial block.
%The initial pebble is put in front of the
%lastly placed $n$th pebble which, having
%done its duty as springboard for this
%block, is subsequently
%irreversibly erased. Therefore, the advancement of
%each block requires two erasures, except the first
%block which requires one, yielding a total
%of $2m-1$ erasures. Let $G = \{ 1,2, \ldots, T_G \}$
%be segmented as $B_1 b_1 \ldots B_m b_m$,
%where each $B_i$ is a copy of interval $I_{n-1}$
%above and $b_i$ is the node following $B_i$, for
%$i=1, \ldots , m$. Hence, $T_G=m 2^{n-1}$.
We give the precise procedure in 
self-explanatory pseudo PASCAL using the procedures
given in section~\ref{reachable}.
%the proof of Lemma~\ref{lem.bennett}.\\

\noindent
\begin{tabbing}
{\bf Procedure}  $A(n,m,G)$: \\ 
 {\bf for} \= $i:=0,1,2, \ldots,m-1$: \\
 \> pebble($i2^n, n$); \\
 \> put springboard on node  $(i+1)2^n$ ; \\
 \> unpebble($i2^n, n$); \\
 \> if $i<m-1$ erase springboard on node $i2^n$ ; \\
\end{tabbing}

The simulation time $T'_G$ is 
$T'_G \approx 2m\cdot 3^{n-1} +2 
\approx 2m ( T_G/m)^{\log 3} = 2m^{1- \log 3 } T_G^{\log 3}$
for $T_G = m2^{n-1}$.
\end{proof}

\begin{theorem}[Space-Irreversibility Trade-off]
\label{theo.si}
(i) Pebble games $G$ of size $2^n-1$ can be won using $n$ pebbles
but not using $n-1$ pebbles.

(ii) If $G$ is a pebble game with a winning strategy
using $n$ pebbles without
erasures, then there is also a winning strategy for $G$
using $E$ erasures and $n-\log (E+1)$ pebbles (for $E$ is an odd
integer at least 1).

\end{theorem}
\begin{proof}
(i) By Corollory~\ref{lem.bennett}.

(ii) By (i), $T_G = 2^n-1$
is the maximum length of a pebble game $G$
for which there is a winning strategy using $n$
pebbles and no erasures.
By Lemma~\ref{lem.erasure}, we can pebble a game $G$
of length $T_G= m2^{n-\log m}=2^n$ using $n+1-\log m$
pebbles and $2m-1$ erasures.
\end{proof}

We analyze the consequences of Theorem~\ref{theo.si}.
 It is convenient
to consider the special sequence of values
$E :=2^{k+2}-1$ for $k:=0,1, \ldots$.
Let $G$ be Bennett's pebble game of Lemma~\ref{lem.bennett}
of length $T_G=2^{n}-1$. 
It can be won using $n$ pebbles
without erasures, or using
$n-k $ pebbles plus $2^{k+2}-1$ erasures (which gives a gain
over not erasing as in Lemma~\ref{lem.bennett} only for $k \geq 1$), but not
using $n-1$ pebbles. 

Therefore, we can exchange space use
for irreversible erasures.
Such a trade-off can be used to reduce
the excessive space requirements of the reversible simulation.
The correspondence between the
erasure pebble game and the
otherwise reversible computations
using irreversible erasures 
that if the pebble game uses $n-k$ pebbles
and $2^{k+2} -1$ erasures, then the otherwise reversible
computation uses $(n-k)S$ space and erases $(2^{k+2}-1)S$ bits
irreversibly. 

Therefore, a reversible simulation of an irreversible
computation of length $T=(2^n-1)S$ can be done using
$nS$ space using $(T/S)^{\log 3 } S$ time,
 but is impossible using $(n-1)S$ space. It can also
be performed using $(n-k)S$ space, $(2^{k+2}-1)S$
irreversible bit erasures and 
 $2^{(k+1)(1-\log 3 )+1} (T/S)^{\log 3} S$
time. In the extreme case
we use no space to store the history and erase about $4T$
bits. This corresponds to the fact that an irreversible
computation may overwrite its scanned symbol irreversibly
at each step.

\begin{definition}
\rm
Consider a simulation 
using $S'$ storage space
and $T'$ time 
which computes 
$y = \langle x, f(x) \rangle$ from $x$ in order to 
simulate
an irreversible computation
using $S$ storage space and $T$ time 
which computes $f(x)$ from $x$.
The {\em irreversible simulation
cost} $B^{S'}(x, y)$ of the simulation
is the number of
irreversibly erased bits in the simulation (with the parameters $S,T,T'$
understood).
\end{definition}

If the irreversible 
simulated computation from $x$ to $f(x)$ uses $T$ steps, then for $S' = nS$ 
and $n =  \log (T/S)$ we have above treated the most space
parsimonious simulation which yields $B^{S'} (x,y) = 0$, with
$y= \langle x,f(x) \rangle$.

\begin{corollary}[Space-Irreversibility Trade-off]
Simulating a $T=(2^{n}-1)S$ step
irreversible computation from $x$ to $f(x)$ 
using $S$ space
 by
a computation from $x$ to $y = \langle x, f(x) \rangle$, the
irreversible simulation cost satisfies:

(i) $B^{(n- k)S } (x,y) \leq B^{nS}(x, y) + (2^{k+2}-1)S$,
for $n \geq k \geq 1$. 

(ii) $B^{(n-1)S}(x,y) > B^{nS}(x,y)$, for $n \geq 1$.

\end{corollary}

For the most space parsimonious
simulation with $n=\log (T/S)$ this means that
\[ B^{S(\log (T/S) - k) } (x,y) \leq 
B^{S \log (T/S)}(x, y) + (2^{k+2}-1)S.\]

\section{Local Irreversible Actions}
Suppose we have an otherwise reversible computation containing
local irreversible actions.
In \cite{LiVi96} it is shown that we can always
simulate such a computation with an otherwise reversible
computation with all irreversibly provided bits
provided at the beginning of the computation, and all
irreversibly erased bits erased at the end of the computation.
This is when we are in the situation when there are no a priori
bounds on the resources
in time or space consumed by the computation.

However, in the case above were there are very tight bounds on the
space used by the computation, we found in Lemma~\ref{lem.erasure}
 a method were at the
cost of limited erasing, precisely controled with
respect to its spacing in the computation
time, we could save on the auxiliary space use.
By Corollary~\ref{lem.pebble} it is {\em impossible} to
shift these erasures to the end of the computation, since 
if we do, then the same
auxiliary space is still needed at precise times spaced during
the simulation time.

Quantum computing is a particular form of reversible computation.
Apart from classical irreversible erasures, quantum computing
has a nonclassical form of irreversibility, namely the 
irreversible observations. An {\em irreversible observation}
makes the superposition of the quantum state of the computer 
collapse from the original state space to a subspace thereof,
where the probability amplitudes of constituent elements
of the new superposition are renormalized.
It is well-known and observed in some papers \cite{Sh94},
that we can replace all observations during the quantum computation by
a composition of observations at the end of the computation.
One wonders if this non-classical type of irreversibility
constituted by irreversible observation of quantum states
also is constrained to strictly local instants during the computation
by restrictions on time or space resources. This seems
to be the case in the $\sqrt{n}$ data item queries 
unstructured database search
algorithm of Grover~\cite{Gr96}. There, we have to observe and renormalize
at precise time instants during the computation to achieve
the improvement of $O(\sqrt{n})$ data item queries in the quantum
algorithm over the classically required $\Omega (n)$ queries.

\section{Reversible Simulation of Unknown Computing Time}
In the previous analysis we have tacitly assumed that the 
reversible simulator knows in advance the number of steps $T$ taken
by the irreversible computation to be simulated.
Indeed, the exhibited  programs $F(\cdot)$ and $A(\cdot)$ have parameters
$I_k$ and $G$  involving $T$. 
In this context one can distinguish on-line computations
and off-line computations to be simulated. On-line computations
are computations which interact with the outside environment
and in principle keep running forever. An example
is the operating system of a computer.
Off-line computations are computations which compute a definite
function from an input (argument) to an output (value). For example,
given as input a positive integer number, compute as output all
its prime factors. For each input such an algorithm will have a definite
running time.

There is a simple device
to remove this dependency for batch computations
 without increasing the simulation time
too much. Suppose we want to simulate a computation with
unknown computation time $T$. Then we simulate $t$ steps
of the computation with $t$ running through the sequence
of values $2,2^{2},2^{3}, \ldots$ For each value $t$ takes on
we reversibly simulate the first $t$ steps of
the irreversible computation. If $T>t$ then the computation is not finished
at the end of this simulation. Subsequently we reversibly undo
the computation until the initial state is reached again, set $t:=2t$
and reversibly simulate again. This way we continue until $t\geq T$
at which bound the computation finishes. The total time spent
in this simulation is
\begin{eqnarray*}
T' & \leq & 2 \sum_{i=1}^{\lceil \log T \rceil} 2^{i \log 3} \\
& \leq & 2 (4T)^{\log 3} .
\end{eqnarray*}

This is the canonical case. With these figures,
just like the original simulation, by suitable choice of parameter $k$
we can obtain $T'=\Theta (T^{1+\epsilon}/S^{\epsilon})$.
for each constant $\epsilon > 0$.

\section*{Acknowledgment}
Wim van Dam pointed at a small error
in the original proof of Lemma~\ref{lem.erasure}
in \cite{LiVi96}. Correcting the proof gives almost the same
result (Lemma~\ref{lem.erasure} in the current paper).

\bibliographystyle{plain}

\end{document}